\documentstyle[12pt]{article} 
\begin{document}

\baselineskip=14pt plus 0.2pt minus 0.2pt
\lineskip=14pt plus 0.2pt minus 0.2pt


\begin{center}
{\bf  
Reply to Comment on \\
``Generating Functions for Hermite Polynomials of Arbitrary Order"}
 
\vspace{0.25in}

\bigskip

Michael Martin Nieto\footnote{\noindent  Email: mmn@pion.lanl.gov}\\
{\it Theoretical Division (MS-B285), Los Alamos National Laboratory\\
University of California \\ 
Los Alamos, New Mexico 87545, U.S.A. \\}
 
\vspace{0.25in}

 D. Rodney Truax\footnote{Email:  truax@acs.ucalgary.ca}\\
{\it Department of Chemistry, University of Calgary\\
Calgary, Alberta T2N 1N4, Canada\\}

\vspace{0.3in}

{Abstract}
\end{center}


The results in the preceding comment are placed on a more 
general mathematical foundation. 

\vspace{0.3in}

\baselineskip=.33in

In  the preceding comment \cite{fer}, our previous results on 
arbitrary-order Hermite generating functions \cite{genf} were 
duplicated and extended.  This was done by using a power-series expansion of
the operator $W=\exp[-\partial^2/4]$ to define the Hermite 
polynomials as $H_n(x)= 2^n W x^n$.  
We observe that this operator definition can be put on a more 
rigorous and general foundation, which allows a better understanding of 
any extended results.

Consider the operator $W(c) \equiv \exp[c\partial^2]$.  It is 
known \cite{miller} that this operator, acting on a well-enough behaved 
function $h(x)$, has the property
\begin{equation}
W(c)h(x) = \frac{1}{\sqrt{4 \pi c}} \int^{\infty}_{-\infty}
             dy \exp\left[-\frac{(y-x)^2}{4c}\right] h(y). \label{mill}
\end{equation}
Then, using $c=-1/4$ in Eq. (\ref{mill}) means that the right-hand integral
(after a change of variable and deformation of the contour of integration)
is a standard integral representation of the Hermite polynomials 
\cite{mos}.  

Even more illuminating is the recognition that the right-hand side of 
Eq. (\ref{mill}) is a Gauss transform of parameter $u=2c$ \cite{bateman}:
\begin{equation}
{\cal G}^{u=2c}_x[h(y)] = W(c)h(x).
\end{equation}
\noindent From this follow the Gauss transforms of interest to us:
\begin{equation}
{\cal G}^{1/2}_x[H_n(y)]= (2x)^n,~~~~~
{\cal G}^{1/2}_x[y^n]=(2i)^{-n}H_n(ix).
\end{equation}
One sees that the necessary condition for obtaining more general 
analytic generating functions  is that an analytic Gauss transform 
can be found. 

Finally, we observe that since even and odd coherent states have 
now been observed in an ion trap \cite{wineland}, it is to be 
hoped that the higher-order coherent states discussed in Ref. \cite{genf}
may also be produced.

This work was supported by the US Department of Energy 
(MMN) and the Natural Sciences and Engineering Research
Council of Canada (DRT).
   


\begin{thebibliography}{99}

\bibitem{fer} F. M. Fern\'andez, Phys. Lett. A (previous comment).

\bibitem{genf} M. M. Nieto and D. R. Truax, 
Phys. Lett. A  208 (1995) 8.

\bibitem{miller}  E. G. Kalnins and W. Miller Jr.,  J. Math. 
Phys. 15 (1974) 1728, Eq. (3.8).

\bibitem{mos} W. Magnus, F. Oberhettinger, and R. P. Soni, {\it 
Formulas and Theorems for the Special Functions of Mathematical Physics}
(Springer, New York, 1966), p. 254.

\bibitem{bateman} A. Erd\'elyi, W. Magnus, F. Oberhettinger, and 
F. G. Tricomi, {\it Higher Transcendental Functions, Vol. II} - Bateman 
Manuscript (McGraw-Hill, New York, 1953), p. 195.

\bibitem{wineland}  C. Monroe, D. M. Meekhof, B. E. King, and 
D. J. Wineland, Science 272 (1996) 1131.

\end{thebibliography}
\end{document}